\documentclass[preprint,showpacs,aps,draft]{revtex4}
\usepackage{epsf}
\usepackage{dcolumn}
\usepackage{bm}
\everymath{\displaystyle}
\begin{document}

\title{Equivalence principle and experimental tests of gravitational spin effects}

\author{\firstname{Alexander J.}~\surname{Silenko}}
\email{silenko@inp.minsk.by}
\affiliation{Institute of Nuclear
Problems, Belarusian State University, Minsk 220080, Belarus}

\author{\firstname{Oleg V.}~\surname{Teryaev}}
\email{teryaev@thsun1.jinr.ru} \affiliation{Bogoliubov Laboratory
of Theoretical Physics, Joint Institute for Nuclear Research,
Dubna 141980, Russia}

\date{\today}

\begin {abstract}
We study the possibility of experimental testing the
manifestations of equivalence principle in spin-gravity
interactions. We reconsider the earlier experimental data and get
the first experimental bound on anomalous gravitomagnetic moment.
The spin coupling to the Earth's rotation may also be explored at
the extensions of neutron EDM and g$-$2 experiments. The spin
coupling to the terrestrial gravity produces a considerable effect
which may be discovered at the planned deuteron EDM experiment.
The Earth's rotation should also be taken into account in optical
experiments on a search for axionlike particles.

\end{abstract}

\pacs {04.20.Cv, 04.25.Nx, 04.80.Cc}
\maketitle

\section{Introduction}

Equivalence principle is known to be one of the basic postulates
of the modern physics, constituting the cornerstone of General Relativity.
Its simplest and well-known
counterpart corresponds to the equality of inertial and gravitational mass
and is tested with good accuracy. The
equivalence principle is also manifested in the interaction
of spin with gravity, as it was first shown in the seminal
paper of I. Yu. Kobzarev and L. B. Okun
\cite{KO}. It means the
absence of both the anomalous gravitomagnetic moment
(AGM) and the gravitoelectric dipole moment which are
gravitational analogs of the anomalous magnetic moment
and the electric dipole moment (EDM), respectively.
It may be derived as a low energy theorem due to the conservation of momentum and orbital angular momentum \cite{KZ}.

Relations obtained by Kobzarev and Okun predict equal frequencies of precession of
quantum (spin) and classical (orbital) angular momenta 
%
and the preservation of helicity of Dirac particles in
gravitomagnetic fields (i.e., the fields defined by the components $g_{i0}$
of the metric tensor,
see Ref. \cite{T1} and references
therein).

This holds in any reference frame in which such components of metric tensor
appear. In particular, one should mention the inertial frame (where
the gravitomagnetic field may be created by the rotation of massive body)
and the rotating noninertial frame.
These properties of spin-gravity interaction
were explored in the number of theoretical papers and suggestions
for experiments (see Refs. \cite{PRD,Mashhoon2,PK} and references therein).
There are also some evidences supporting the
conjecture \cite{T2} that the absence of the AGM
is valid separately for quarks and
gluons in the nucleon, which may be related to the phenomena of
confinement and spontaneous chiral symmetry breaking.

The experimental tests of the Kobzarev-Okun relations are lacking and are
therefore of much importance. 
The problem of existence of
the dipole
spin-gravity coupling in a static gravitational field
has been
discussed for a long time (see Refs. \cite{PRD,O,Mashhoon2} and
references therein). 
Evidently, this 
coupling given in the form of $\bm S\cdot\bm g$ ($\bm g$ is the
acceleration) contradicts to the theory \cite{PRD,PK} and
violates both the CP invariance and the relation predicting the
absence of the gravitoelectric dipole moment \cite{KO}.
Therefore, the
negative result of realized experiments
imposes some restrictions not only on the spin-gravity coupling but
also on the gravitoelectric dipole moment.

In this article we carefully reanalyze the results of spin experiments with
Hg atoms and get the first experimental bound on their AGMs.
We also
suggest to 
extend some experiments
with spinning particles 
for testing the absence of the AGMs
and calculate
related gravitational effects.

\section{Spin coupling to gravity and rotation}

Spin rotation due to the action of the Earth's gravity is
\cite{PK,PRD}
\begin{eqnarray}
\frac{d\bm S}{dt}=\bm\Omega_g\times\bm S, ~~~
\bm\Omega_g=-\frac{2\gamma
+1}{\gamma (\gamma +1)mc^2} \bm g\times\bm
p, \label{eq7}\end{eqnarray} where $\bm S$
is the spin vector and $\gamma$ is the Lorentz factor.
In deriving this equation, we neglected small corrections due to the
derivatives from the metric tensor including those depending on curvature.
The maximum value of $\Omega_g$ is $2g/c$. 

We have carried out the relativistic generalization of the pioneering results \cite{HN}
on the effect of Earth's rotation on particle spin.
We have performed the {\it exact} Foldy-Wouthuysen
transformation (see Ref. \cite{JMP}) of Dirac Hamiltonian \cite{HN}.
For the particle in the rotating frame, this Hamiltonian takes the form
($\hbar=c=1$)
\begin{equation} {\cal H}=\beta m+{\cal E}+{\cal
O},
\label{eqh}\end{equation} where
\begin{equation}\begin{array}{c}
{\cal E}=-\bm\omega\cdot\bm J, ~~~ \bm J=\bm L+\bm S, ~~~
{\cal O}=\bm\alpha \cdot\bm p.
\end{array}\label{eqh1}\end{equation}
$\bm\omega$ is the
angular frequency of the Earth's rotation,
$\bm L=\bm r\times\bm p$ and $\bm S=\bm\Sigma/2$ are the angular
momentum and spin operators, $\bm
p=-i\nabla$ is the momentum operator, ${\cal E}$ and ${\cal O}$ mean
even and odd terms that commute and anticommute with the matrix
$\beta$, respectively. We use commonly accepted definitions of Dirac
matrices \cite{FW}.

The Hamiltonian in the Foldy-Wouthuysen representation is given by
\begin{eqnarray}
{\cal H}_{FW}=\beta\sqrt{m^2+\bm p^2}-\bm\omega\cdot\bm J.  
\label{eqn}\end{eqnarray}
and the lower spinor in this representation is zero.

It is very important that exact Hamiltonian (\ref{eqn}) does not
contain any quantum corrections to the classical Hamiltonian derived
by Mashhoon \cite{Mashhoon} and that is also agrees with the earlier result
by Gorbatsevich \cite{Gor}. The equal coupling of rotation to orbital
and spin momenta (which is not true 
for a magnetic field) is a manifestation of the absence of the AGMs.

The particle motion is
characterized by the operators of velocity and acceleration:
\begin{eqnarray}
v^i\equiv\frac{dx^i}{dx^0}=i[{\cal H},x^i],~~~ x^0\equiv t,\nonumber\\
w^i\equiv\frac{d v^i}{dx^0}=i[{\cal H},v^i]=-\left[{\cal H},[{\cal
H},x^i]\right]. \label{eq19}\end{eqnarray}
In the considered case, these operators are equal to
\begin{eqnarray}
\bm v=\beta\frac{\bm p}{\epsilon}-\bm\omega\times\bm r, ~~~
\epsilon =\sqrt{m^2+\bm p^2},
\nonumber\\
\bm w=2\beta\frac{\bm p\times\bm\omega}{\epsilon}
+\bm\omega\times(\bm\omega\times\bm r)\nonumber\\=2\bm
v\times\bm\omega-\bm\omega\times(\bm\omega\times\bm r). \label{eqvn}\end{eqnarray}

Eq. (\ref{eqvn}) also results in the quantum formula for
the force acting on the relativistic particle which coincides with the
classical formula \cite{GG} for
the sum of the Coriolis and centrifugal forces.
Thus, the classical and quantum approaches are in full agreement.




\section{Experiments with atoms and cold neutrons}

AGM would manifest itself in the coupling to any (nonflat) metric, in
particular, in the case of rotating frames.
In this section, we discuss
the emerging opportunity to test the absence of the AGMs provided by experiments
with atoms and cold neutrons. The experiments are performed with two
kinds of atoms (or with neutrons and atoms) designated by indices 1
and 2.

Let us reconsider the earlier results \cite{Venema} as restrictions on
the AGM rather than on the
dipole spin-gravity coupling. Recall that latter violates not only
the Kobzarev-Okun relation for the gravitoelectric dipole moment
but also CP invariance and may be neglected.
The spin-dependent Hamiltonian for atoms in $S$ states
may be obtained by the modification of the coefficient of
the term defining the spin-rotation coupling
and has
the form
\begin{eqnarray}
{\cal H}=-g\mu_N\bm B\cdot\bm S-\zeta\hbar\bm\omega\cdot\bm S,~~~
\zeta=1+\chi, \label{anml}
\end{eqnarray} where $g$ is the nuclear $g$ factor, $\mu_N$ is the nuclear magneton,
and $\chi$ is the AGM.
The measured ratio of energy differences in neighboring Zeeman levels,
$R=|\nu_{2}|/|\nu_{1}|$, depends on the AGMs. The difference
of these ratios for two opposite directions of magnetic
field 
is given by
\begin{eqnarray} R_+-R_-=\pm\frac{2f \cos {\theta}}{|\nu_1|}\left(
\zeta_{2}-{\cal G}\zeta_{1}\right),~~~{\cal G}=\frac{g_2}{g_1},
\label{sol}
\end{eqnarray} where $\theta$ is the angle between the directions of magnetic
field and the Earth's rotation axis, $f=\omega/(2\pi)=11.6~\mu$Hz is
the Earth's rotation frequency, and $|\nu_1|$ is the Zeeman
frequency for atoms of the first kind.
The experimental conditions of
\cite{Venema} for $^{199}$Hg and $^{201}$Hg atoms correspond to
$\theta\approx0,~{\cal G}=-0.369139$.
Reconsidering the bound for $R_+ - R_-$
obtained in that Ref., we drop the contribution of
CP-violating gravitoelectric dipole moment, but account for the
possibility for nonzero AGM, which makes a difference between (\ref{sol})
and their Eq. (4).
As a result, their data lead to the following
restriction:
$$|\chi(^{201}{\rm Hg})+0.369\chi(^{199}{\rm Hg})|<0.042 ~~~(95\% {\rm C.L.}).$$
To our best knowledge, this is the first experimental bound on the AGM,
and consequently the first test of the Kobzarev-Okun relations.
The sensitivities of similar experiments fulfilled with deuterium
\cite{WR} and beryllium \cite{Wl} atoms are not sufficient to obtain
significant restrictions.

Another experiment is fulfilled at Institute Laue-Langevin (ILL) with
ultracold neutrons placed in electric and magnetic fields \cite{baker}
and aimed to search for their EDM. There is a
recent claim \cite{com} that spin-rotation coupling should be
already taken into account when analyzing the data obtained.
To address the problem of testing the absence of the AGMs, 
the data for the opposite directions of magnetic 
field should be considered separately, while averaging 
over the directions of electric field should be performed. The
correction for the Earth's rotation is rather large and corresponds
to the EDM of $1.7\times10^{-24}$ e$\cdot$cm when $E=10$ kV/cm.
%
The expected sensitivity of this experiment to the AGM is also of
order of $10^{-2}$.
It is also possible \cite{Anandan} to use the magnetic resonance methods
for atomic and molecular beams.

Spin coupling to the Earth's rotation may in principle also be
investigated in the GRANIT (GRAvitational Neutron Induced Transitions) \cite{granit1,granit2} experiment, where
quantum states of cold neutrons in the terrestrial gravitational
field were observed.


Ultracold neutrons can also be used in interferometer experiments
with rotating spin-flippers \cite{Mashexp} and implemented at the
existing and developed interferometers at ILL and Tokai
\cite{Private}. It seems reasonable to have two (rather than one as
suggested in \cite{Mashexp}) rotating spin-flippers.
Signals should be absent if they are rotated in the same directions.
In the case of rotation in opposite directions, signals should be
twice larger in comparison when only one flipper rotates.



\section{Comparison of Spin-Rotation Coupling for Electrons and Positrons}


The above mentioned experiments do not solve the important problem
of the equivalence of gravitational effects for particles and
antiparticles which may be tested in the storage rings.
The corresponding equation of spin motion in a cylindrical coordinate system
\cite{PhysRevST} with an addition of the
gravitational correction is given by
\begin{eqnarray}
\frac{d\bm S}{dt}=\bm\omega_{a}\times\bm S, ~~~
\bm\omega_{a}=\bm\Omega+\bm\Omega_{EDM}
+\bm\omega+\bm\Omega_g,
\label{eqc}\end{eqnarray}
%
where $\bm\Omega$ and $\bm\Omega_{EDM}$ are angular velocities of
spin rotation caused by magnetic and electric dipole moments,
respectively. The (pseudo)vectors $\bm\Omega$ and $\bm\Omega_{EDM}$
are oppositely directed for particles and antiparticles.
The gravitational corrections to the angular velocity of spin
rotation in Cartesian and cylindrical coordinates are
$-\bm\omega+\bm\Omega_g$ and $\bm\omega+\bm\Omega_g$, respectively.




It is the quantity $\omega_{a}$ which is measured in storage ring and
Penning trap experiments. The Earth's rotation can simulate the CPT
violation because it brings a fictitious difference between $g$
factors of electron and positron. The measurements of electron and
positron $g$ factors in the Penning trap at the level of accuracy
of order of 0.1 Hz \cite{CPT,CPT2} were not sensitive to the
Earth's rotation.


To make the gravitational corrections 
observable, it is
desirable to use a relatively weak magnetic field in order to
decrease the spin rotation frequency as much as possible. Since this
frequency is proportional to the cyclotron one, the particle
trajectory should be extended and gravitational experiments should
be performed in storage rings. 
 is equal

The best condition for the comparison of the spin-rotation coupling
for 
electrons and positrons 
is perhaps provided by the use of a muon g$-$2 ring (namely, the
7.11 m ring of the Brookhaven National Laboratory). The electron/positron beam
polarization may be measured with the methods described in Refs. \cite{Sinc,GE}.
The frequency of spin rotation (g$-$2 frequency) actually measured
with the accuracy of 0.16 Hz (0.7 ppm) \cite{fin} is almost the same
for muons and electrons/positrons. 
The best 
sensitivity of experiment with electrons and positrons
can be achieved with electric focusing and the ``magic" Lorentz
factor ensuring 
a dramatic reduction of the influence of the electric
field on the spin rotation  
%
and resulting in a small width of resonance line \cite{FS,fin}.
%


The sensitivity of the proposed experiment is not affected by the
systematical error 
in measurement of the magnetic field because it is the same for
electrons and positrons and therefore is canceled in the difference
$\omega_a(e^+)-\omega_a(e^-)$.
To compare the sensitivities of the proposed experiment and the muon
g$-$2 experiment, one should take into consideration only systematical
errors due to the electric field and the fitting procedure. The
systematical errors caused by the horizontal and vertical coherent
betatron oscillations (0.07 and 0.04 ppm, respectively, for the
muons \cite{fin}) are much less important for the electrons and
positrons because the decay time of these oscillations ($\sim
100~\mu$s \cite{fin}) is very small in comparison with the beam
circulation time. These systematical errors can additionally be
reduced due to the fact that the electric focusing is 207 times
stronger for the electrons/positrons than for the muons.
The shift of 
the precession frequency due to the electric
field depends on the momentum spread and is given by (Eqs. (17)
and (21) in Ref. \cite{fin})
\begin{eqnarray}
\frac{\delta\omega_a}{\omega_a}=-2\beta^2\frac{n}{1-n}
\left(\frac{p-p_0}{p_0}\right)^2, \label{momsp}\end{eqnarray}
where $n$ is the field index and $\beta=v/c$. The momentum spread, $(p-p_0)/p_0$,
equal to 0.5$\%$ for the muons \cite{fin} can be considerably less
(right up to $10^{-6}$ \cite{prcom}) for the electron and positron
beams providing a great reduction of the systematical error.
In the muon g$-$2 experiment, this systematical error was
$\sim$0.01 Hz \cite{fin}, i.e., about 10$\%$ of the electric field
correction and about 1$\%$ of the linewidth
$\sqrt{<(\delta\omega_a)^2>}$. 
We suppose that a 
relation between these
quantities 
cannot be very different in the proposed experiment.
If the momentum spread of the electrons/positrons is
$5\times10^{-5}$ (two orders of magnitude less than for the muons)
and $n\leq0.2$, the linewidth is reduced $10^{4}$ times in
comparison with the muon g$-$2 experiment. Even if the related
systematical error would be $6\%\div8\%$ of the linewidth, the resulting error of frequency
determination is about $10~\mu$Hz.
Besides the comparison of gravitational spin-rotation coupling for particles and antiparticles, the restriction on the CPT
violation would also be improved. 

\section{Effect of Earth's Gravity on Spin Dynamics in Storage Rings}

To measure the effect of the Earth's gravity on spin dynamics, one
needs to detect the spin rotation about a horizontal axis. 
The detection
can 
be provided 
if the particle spin is governed by a uniform upward magnetic field and
a resonant longitudinal electric one ($\bm E\|\bm v$).
This field configuration corresponds to the resonant deuteron
electric-dipole-moment (dEDM) experiment \cite{OMS}.

When magnetic focusing is used, the gravitational force acting on
particles, $\bm F_g=(2\gamma^2-1)m\bm g/\gamma$,
defines the nonzero radial magnetic field which
causes the spin turn with the average angular velocity
\begin{eqnarray}
\bm\omega_{m}=\frac{(1+a\gamma)(2\gamma^2-1)}{mc^2\gamma(\gamma^2-1)}
\bm g\times\bm p. \label{eql}\end{eqnarray} The resulting angular
velocity $\bm\omega_{a}$ has the vertical and radial components
and is given by
\begin{eqnarray}
\bm\omega_{a}=\Omega_z\bm e_z+\bm\Omega_{EDM}
+\bm\Omega_g+\bm\omega_{m}, \label{eqcnw}\end{eqnarray}
while the average radial component of angular velocity of Earth's rotation  is zero. If
we disregard terms describing systematical errors, 
Eq. (\ref{eqcnw}) takes the form (here and below $\hbar=c=1$)
\begin{eqnarray}
\bm\omega_{a}=-\frac{ea}{m}B_z\bm
e_z-\frac{d}{S}\left(\frac{1}{\gamma}\bm E +\bm\beta\times\bm
B\right)\nonumber\\+\frac{\left[1+a(2\gamma^2-1)\right]\gamma}{\gamma^2-1}
\bm g\times\bm\beta, \label{eqd}\end{eqnarray} where $d$ is the
EDM and $S$ is the spin quantum number.

In Eq. (\ref{eqd}), the quantities $\bm E$ and $\bm\beta$
oscillate at a near-resonant frequency (see Ref. \cite{preprint}).
The resulting buildup of the vertical polarization
calculated by the method elaborated in Ref. \cite{preprint}
is equal to
\begin{eqnarray}
P_z=-\frac{1}{2}P_0\Delta\beta_m\sin{(\psi-\varphi_m)}\biggl\{
\frac{d}{S}B_0\left(1+a\gamma^2\right) \nonumber\\ +
g|\sin{\Phi}|\frac{\gamma^3}{\gamma^2-1}\left[1-a(2\gamma^2-3)\right]\biggr\}
t, \label{eqdel}\end{eqnarray} where $\Delta\beta_m\equiv \Delta
v_m/c$ and $\varphi_m$ 
characterize the resonant modulation of the beam velocity 
\cite{OMS}, $\psi$ is the azimuthal angle of spin direction (with respect to the
$\bm e_\rho$ axis) at
zero time, $\Phi$ is the geographic latitude, and $P_0$ is the 
polarization of the incident beam.
In the planned dEDM experiment, the Earth's gravity would bring
the effect identical to that given by the deuteron EDM of
$d=2\times10^{-29}$ e$\cdot$cm. This effect is rather important,
because the expected sensitivity of the dEDM experiment \cite{OMS} is of the same order.

It is rather difficult to incorporate the AGM $\chi$ into Eqs.
(\ref{eqd}),(\ref{eqdel}) in the general case because the initial Dirac
equation does not contain the AGM. However, the AGM can be inserted into
Eq. (\ref{eq7}) in the nonrelativistic approximation when one keeps
quantities of order of $\beta$ and neglects those of order of $\beta^2$.
In this approximation,
the angular velocity of spin rotation, $\bm\Omega_g$,
should be proportional
to the total gravitomagnetic moment and
may be obtained by the modification of the respective coefficient:
$$
\bm\Omega_g=-\frac{3\zeta}{2c} \bm g\times\bm
\beta.
$$
As a result, the quantity $\chi\bm\Omega_g/\zeta$ should be added to the
right-hand side of Eq. (\ref{eqd}). Other terms in Eq. (\ref{eqcnw}) are
not affected by the AGM and Eq. (\ref{eqd}) takes the form ($c=1$)
\begin{eqnarray}
\bm\omega_{a}=-\frac{ea}{m}B_z\bm
e_z-\frac{d}{S}\left(\frac{1}{\gamma}\bm E +\bm\beta\times\bm
B\right)\nonumber\\+\left\{\frac{\left[1+a(2\gamma^2-1)\right]\gamma}{\gamma^2-1}
-\frac{3\chi}{2}\right\}
\bm g\times\bm\beta. \label{eqdm}\end{eqnarray}

In the planned deuteron EDM experiment, $\gamma=1.28$
\cite{OMS} and such a consideration of the AGM is applicable.
However, one should take into account that magnetic focusing does not
affect a particle at rest and the equality $\beta=0$ results in the
divergence of the quantity $\bm\omega_{m}$.


\section{Optical effects caused by the Earth's rotation}

Photon polarization is significantly influenced
by the Earth's rotation. When frequencies of left-circularly and
right-circularly polarized electromagnetic waves coincide in an
inertial frame, they differ in a
rotating frame
\cite{MashPLA}. This effect has been observed (see Ref.
\cite{Ashby} and references therein). The plane of polarization of
a linearly polarized electromagnetic wave rotates in a stationary
(but nonstatic) spacetime (Skrotskii effect \cite{Skrotskii}).
This effect results in an optical rotation of electromagnetic wave
in vacuum caused by the Earth's rotation and defined by
$d\phi/dl=\bm\omega\cdot\bm l_0/c$ \cite{KopeMash}, where $\bm
l_0$ is the unit vector pointing in the wave direction and
$\omega/c=2.43\times10^{-10}$ rad/km. This relatively large 
optical rotation has not been taken into consideration in the
Brookhaven, Fermilab, Rochester, Trieste (BFRT)
\cite{BFRT} and PVLAS \cite{PVLAS} experiments on a search for
axionlike particles.

In the PVLAS experiment, the light direction is vertical,
$\bm\omega\cdot\bm l_0=\omega\sin{\Phi}$, and the effect of the
Earth's rotation is
$d\phi/dl=1.73\times10^{-10}$ rad/km. 
This value corresponds to the optical rotation
$\alpha_E=1.1\times10^{-12}$ rad/pass and therefore is of the same
order as the much discussed effect observed by the PVLAS
collaboration: $\alpha = (3.9\pm0.5)\times10^{-12}$ rad/pass
\cite{PVLAS}. Evidently, the Skrotskii effect can be discovered in
the framework of the PVLAS experiment and it can be used for
checking the sensitivity. It can also be measured in a similar
experiment performed without magnetic field.

The effect of the Earth's rotation did not become apparent in the
BFRT experiment because all effects independent of the angle
between the plane of
polarization and the magnetic field direction were eliminated \cite{BFRT}. 

\section{Conclusions}

There is a number of 
possibilities to measure the coupling of spin to rotation and
gravity and therefore to verify the Kobzarev-Okun relations.
We suggest the
reinterpretation of earlier experiment with atomic spin
\cite{Venema} leading to the first check of the AGMs at few per cent level of
accuracy.
The straightforward extensions of experiments with (ultra)cold
neutrons can also provide the important test of the absence of the AGMs.
Possible gravitational experiment in the g$-$2 ring 
enables to compare the spin-rotation coupling for particles
(electrons) and antiparticles (positrons). The proposed extension
of the deuteron EDM experiment gives an exciting opportunity to
detect the spin-gravity coupling existing only for moving
particles. The Earth's rotation should be taken into account in
optical experiments on a search for axionlike particles, where
observed  effect \cite{PVLAS} is of the same order as that of the
Earth's rotation.

\section*{Acknowledgements}

This work was supported in part by the Belarusian
Republican Foundation for Fundamental Research (Grant
No. F06D-002), the Deutsche Forschungsgemeinschaft
(Grant No. 436 RUS 113/881/0), the Russian Foundation
for Basic Research (Grant No. 03-02-16816), and the
Russian Federation Ministry of Education and Science
(Grant No. MIREA 2.2.2.2.6546). We have been informed
that another group intends to make an estimate of the level
of perturbation on several precision experiments involving
spin \cite{Priv}.
We are grateful to C.J.G. Onderwater for this information and helpful correspondence.
We are also indebted to G. Papini for interest in our work and bringing
to our attention Refs. \cite{42,43,44}.
O.T. is thankful to V.~Nesvizhevsky, K.~Protasov, H. Rauch and H. Shimizu
for useful discussions.

\end{document}